\documentclass[reprint,twocolumn,aps,prc,a4paper,superscriptaddress,showpacs,preprintnumbers,amsmath,amssymb]{revtex4}
\usepackage{amssymb}
\usepackage{amsmath}
\usepackage{graphicx}
\usepackage{lscape}
\usepackage{booktabs,longtable}
\usepackage{mathptmx, courier, pifont}
\usepackage[scaled=0.92]{helvet}
\usepackage[T1]{fontenc}
\usepackage{textcomp}

\begin{document}

\title{Application of Isochronous Mass Spectrometry for the study of angular momentum population in projectile fragmentation reactions}
\author{X.~L.~Tu}
\affiliation{Institute of Modern Physics, Chinese Academy of Sciences, Lanzhou 730000, People's Republic of China}
\affiliation{GSI Helmholtzzentrum f\"{u}r Schwerionenforschung, Planckstra{\ss}e 1, 64291 Darmstadt, Germany}
\affiliation{Max-Planck-Institut f\"{u}r Kernphysik, Saupfercheckweg 1, 69117 Heidelberg, Germany}

\author{A.~Keli\'c-Heil}
\affiliation{GSI Helmholtzzentrum f\"{u}r Schwerionenforschung, Planckstra{\ss}e 1, 64291 Darmstadt, Germany}

\author{Yu.~A. Litvinov}
\affiliation{Institute of Modern Physics, Chinese Academy of Sciences, Lanzhou 730000, People's Republic of China}
\affiliation{GSI Helmholtzzentrum f\"{u}r Schwerionenforschung, Planckstra{\ss}e 1, 64291 Darmstadt, Germany}
\affiliation{Max-Planck-Institut f\"{u}r Kernphysik, Saupfercheckweg 1, 69117 Heidelberg, Germany}

\author{Zs.~Podoly\'ak}
\affiliation{Department of Physics, University of Surrey, Guildford GU2 7XH, United Kingdom}

\author{Y.~H.~Zhang}
\affiliation{Institute of Modern Physics, Chinese Academy of Sciences, Lanzhou 730000, People's Republic of China}

\author{W.~J.~Huang}
\affiliation{Institute of Modern Physics, Chinese Academy of Sciences, Lanzhou 730000, People's Republic of China}
\affiliation{Graduate University of Chinese Academy of Sciences, Beijing, 100049, People's Republic of China}
\affiliation{CSNSM, Univ. Paris-Sud, CNRS/IN2P3, Universit\'{e} Paris-Saclay, 91405 Orsay, France}

\author{H.~S.~Xu}
\affiliation{Institute of Modern Physics, Chinese Academy of Sciences, Lanzhou 730000, People's Republic of China}

\author{K.~Blaum}
\affiliation{Max-Planck-Institut f\"{u}r Kernphysik, Saupfercheckweg 1, 69117 Heidelberg, Germany}

\author{F. Bosch}
\affiliation{GSI Helmholtzzentrum f\"{u}r Schwerionenforschung, Planckstra{\ss}e 1, 64291 Darmstadt, Germany}

\author{R.~J.~Chen}
\affiliation{Institute of Modern Physics, Chinese Academy of Sciences, Lanzhou 730000, People's Republic of China}

\author{X.~C.~Chen}
\affiliation{Institute of Modern Physics, Chinese Academy of Sciences, Lanzhou 730000, People's Republic of China}
\affiliation{GSI Helmholtzzentrum f\"{u}r Schwerionenforschung, Planckstra{\ss}e 1, 64291 Darmstadt, Germany}
\affiliation{Graduate University of Chinese Academy of Sciences, Beijing, 100049, People's Republic of China}

\author{C.~Y.~Fu}
\affiliation{Institute of Modern Physics, Chinese Academy of Sciences, Lanzhou 730000, People's Republic of China}

\author{B.~S.~Gao}
\affiliation{Institute of Modern Physics, Chinese Academy of Sciences, Lanzhou 730000, People's Republic of China}
\affiliation{GSI Helmholtzzentrum f\"{u}r Schwerionenforschung, Planckstra{\ss}e 1, 64291 Darmstadt, Germany}
\affiliation{Max-Planck-Institut f\"{u}r Kernphysik, Saupfercheckweg 1, 69117 Heidelberg, Germany}
\affiliation{Graduate University of Chinese Academy of Sciences, Beijing, 100049, People's Republic of China}

\author{Z.~Ge}
\affiliation{Institute of Modern Physics, Chinese Academy of Sciences, Lanzhou 730000, People's Republic of China}

\author{Z.~G.~Hu}
\affiliation{Institute of Modern Physics, Chinese Academy of Sciences, Lanzhou 730000, People's Republic of China}

\author{D.~W.~Liu}
\affiliation{Institute of Modern Physics, Chinese Academy of Sciences, Lanzhou 730000, People's Republic of China}

\author{S.~A.~Litvinov}
\affiliation{GSI Helmholtzzentrum f\"{u}r Schwerionenforschung, Planckstra{\ss}e 1, 64291 Darmstadt, Germany}

\author{X.~W.~Ma}
\affiliation{Institute of Modern Physics, Chinese Academy of Sciences, Lanzhou 730000, People's Republic of China}

\author{R.~S.~Mao}
\affiliation{Institute of Modern Physics, Chinese Academy of Sciences, Lanzhou 730000, People's Republic of China}

\author{B.~Mei}
\affiliation{GSI Helmholtzzentrum f\"{u}r Schwerionenforschung, Planckstra{\ss}e 1, 64291 Darmstadt, Germany}

\author{P.~Shuai}
\affiliation{Institute of Modern Physics, Chinese Academy of Sciences, Lanzhou 730000, People's Republic of China}
\affiliation{University of Science and Technology of China, Hefei 230026, China}

\author{B.~H.~Sun}
\affiliation{School of Physics and Nuclear Energy Engineering, Beihang University, Beijing 100191, People's Republic of China}

\author{Y.~Sun}
\affiliation{Institute of Modern Physics, Chinese Academy of Sciences, Lanzhou 730000, People's Republic of China}
\affiliation{Department of Physics and Astronomy, Shanghai Jiao Tong University, Shanghai 200240, People's Republic of China}

\author{Z.~Y.~Sun}
\affiliation{Institute of Modern Physics, Chinese Academy of Sciences, Lanzhou 730000, People's Republic of China}

\author{P.~M.~Walker}
\affiliation{Department of Physics, University of Surrey, Guildford GU2 7XH, United Kingdom}

\author{M.~Wang}
\affiliation{Institute of Modern Physics, Chinese Academy of Sciences, Lanzhou 730000, People's Republic of China}

\author{N. Winckler}
\affiliation{GSI Helmholtzzentrum f\"{u}r Schwerionenforschung, Planckstra{\ss}e 1, 64291 Darmstadt, Germany}

\author{J.~W.~Xia}
\affiliation{Institute of Modern Physics, Chinese Academy of Sciences, Lanzhou 730000, People's Republic of China}

\author{G.~Q.~Xiao}
\affiliation{Institute of Modern Physics, Chinese Academy of Sciences, Lanzhou 730000, People's Republic of China}

\author{Y.~M.~Xing}
\affiliation{Institute of Modern Physics, Chinese Academy of Sciences, Lanzhou 730000, People's Republic of China}

\author{X.~Xu}
\affiliation{Institute of Modern Physics, Chinese Academy of Sciences, Lanzhou 730000, People's Republic of China}
\affiliation{Graduate University of Chinese Academy of Sciences, Beijing, 100049, People's Republic of China}

\author{T.~Yamaguchi}
\affiliation{Department of Physics, Saitama University, Saitama 338-8570, Japan}

\author{X.~L.~Yan}
\affiliation{Institute of Modern Physics, Chinese Academy of Sciences, Lanzhou 730000, People's Republic of China}
\affiliation{Max-Planck-Institut f\"{u}r Kernphysik, Saupfercheckweg 1, 69117 Heidelberg, Germany}
\affiliation{Graduate University of Chinese Academy of Sciences, Beijing, 100049, People's Republic of China}

\author{J.~C.~Yang}
\affiliation{Institute of Modern Physics, Chinese Academy of Sciences, Lanzhou 730000, People's Republic of China}

\author{Y.~J.~Yuan}
\affiliation{Institute of Modern Physics, Chinese Academy of Sciences, Lanzhou 730000, People's Republic of China}

\author{Q.~Zeng}
\affiliation{Institute of Modern Physics, Chinese Academy of Sciences, Lanzhou 730000, People's Republic of China}
\affiliation{University of Science and Technology of China, Hefei 230026, China}

\author{W.~Zhang}
\affiliation{Institute of Modern Physics, Chinese Academy of Sciences, Lanzhou 730000, People's Republic of China}
\affiliation{Graduate University of Chinese Academy of Sciences, Beijing, 100049, People's Republic of China}

\author{H.~W.~Zhao}
\affiliation{Institute of Modern Physics, Chinese Academy of Sciences, Lanzhou 730000, People's Republic of China}

\author{T.~C.~Zhao}
\affiliation{Institute of Modern Physics, Chinese Academy of Sciences, Lanzhou 730000, People's Republic of China}

\author{X.~H.~Zhou}
\affiliation{Institute of Modern Physics, Chinese Academy of Sciences, Lanzhou 730000, People's Republic of China}

\date{\today}

\begin{abstract}
Isochronous mass spectrometry was applied to measure isomeric yield ratios of fragmentation reaction products.
This approach is complementary to conventional $\gamma$-ray spectroscopy in particular for measuring yield ratios for long-lived isomeric states.
Isomeric yield ratios for the high-spin $I=19/2\hbar$ states in the mirror nuclei $^{53}$Fe and $^{53}$Co are measured to study angular momentum population following the projectile fragmentation of $^{78}$Kr at energies of $\sim$ 480 $A$ MeV on a beryllium target. The 19/2 state isomeric ratios of $^{53}$Fe produced from different projectiles in literature have also been extracted as a function of mass number difference between projectile and fragment (mass loss). The results are compared to  ABRABLA07 model calculations.
The isomeric ratios of $^{53}$Fe produced using different projectiles suggest that the theory underestimates not only the previously reported dependence on the spin but also the dependence on the mass loss.
\end{abstract}


\maketitle

\section{Introduction}
Fragmentation nuclear reactions are often used at intermediate and high projectile energies (typically $50-1500$~$A$ MeV)
to produce exotic nuclei. In the reaction process, some nucleons of the projectile are abraded at the overlapping zone between the target and projectile nuclei.
As a consequence, the remaining projectile is highly excited and promptly de-excites by evaporating nucleons until a final fragment is formed with an excitation energy below the particle emission threshold~\cite{Morrissey98}.
The projectile fragments are characterised by relatively small angular and energy dispersion
and show a very strong kinematical focusing to forward angles.

In combination with an in-flight spectrometer, which takes advantage of the reaction kinematics,
one can separate nuclei of interest on very short time scales down to a few 100 ns~\cite{Geissel02}.
The projectile fragmentation reaction has been proven to be an important tool for producing nuclei far from stability.
It is considered to be one of the main reactions to produce even more exotic systems at the next-generation radioactive beam facilities, as e.g., FAIR at GSI~\cite{Henning08} and HIAF at IMP~\cite{Yang13}, where one of the goals is to investigate nuclides
around the $r$-process path of stellar nucleosynthesis~\cite{Burbidge57}.

The production cross sections ($\sigma$)~\cite{Summerer00}, momentum distributions ($P$)~\cite{Tarasov04}
and angular momenta ($J$)~\cite{Jong97} of the produced fragments are basic properties of the fragmentation reaction.
Their knowledge is indispensable for the understanding of the collision process
and, in turn, is essential for the design and operation of the fragment separator facilities.

Population of a particular excited state with a given angular momentum $J$ can usually not be measured in experiment
due to its prompt de-excitation (< 10 ns)~\cite{Bacelar13}.
However, the production cross sections and momentum distributions of the de-excited nuclei
in a ground or in a long-lived isomeric state are routinely measured.
It is well established that the isomers are produced in the projectile fragmentation reaction~\cite{Pfutzner02}.
The promptly decaying excited states may feed an isomeric state which then survives until the measurement is performed.
By assuming that all excited states with $J$ ($J>J_m$) eventually decay to the isomeric state of interest with $J_m$,
the measured isomeric yield ratio, which is the number of nuclei populated in an isomeric state relative to their total number,
reveals the integrated population probability of the states having the excitation energies $E^*\geq E^*_m$
and angular momenta $J\geq J_m$, where $E^*_m$ indicates the excitation energy of isomer.
Thus an integral population probability of high angular momentum can be deduced from experiment.

Recently, isomeric yield ratios have been obtained from $\gamma$-ray
spectroscopy investigations of angular momentum population of projectile 
fragments~\cite{Bacelar13, Gladnishki04, Pfutzner02, Pfutzner98, Myalski09, Steer11, Bowry13, Myalski12, Podolyak05, Podolyak06}. 
$\gamma$-ray spectroscopy is a powerful technique, which provides valuable contributions to nuclear structure investigations.
However, it is typically limited to studies of excited states with half-lives between
about 100~ns and several milliseconds~\cite{Pfutzner02, Gladnishki04}.
The lower half-life limit is given by the time of flight through a beam line and the upper limit
is determined by the need to correlate delayed $\gamma$-rays with individual ions in a segmented implantation detector with a technologically limited number of pixels~\cite{Gladnishki04}.

The latter limitation can be overcome using other experimental techniques. 
A nuclear isomer has the same number of protons and neutrons as the corresponding 
nuclear ground state but posses an excitation energy, which is reflected in a heavier mass.
The mass difference between isomeric and ground states can directly be resolved by high-resolution mass spectrometry. 
Recently, numerous isomers were investigated with storage-ring \cite{Sun10, Chen13} 
and Penning-trap~\cite{Block08, Roosbroeck04} mass spectrometry.

In the case of a storage ring, the experiments are performed at high energies employing highly-charged ions.
The revolution times $T$ of various stored ions are related (in first order)
to their mass-to-charge ratios $m/q$ via~\cite{Franzke08}
%===============  eq. 1  ========================================
\begin{equation}
\frac{\Delta T}{T}=\frac{1}{\gamma_t^2} \frac{\Delta
(m/q)}{(m/q)}-\left(1-\frac{\gamma^2}{\gamma_t
^2}\right)\frac{\Delta v}{v}\quad,
\label{eq1}
\end{equation}
%=================================================================
where $\gamma $ is the relativistic Lorentz factor and $\gamma_t$ denotes the transition energy of the storage ring.
In order to resolve and to determine $m/q$ values by the revolution times of the ions,
the term containing their velocity spreads, $\Delta v/v$, needs to be made negligibly small.
For this purpose, two complementary experimental techniques have been developed,
namely Schottky (SMS) and Isochronous Mass Spectrometry (IMS)~\cite{Franzke08}.

In the IMS ion-optical mode, the energy of the stored ions of interest is chosen such that $\gamma \approx \gamma_t$.
This leads to the situation that a faster ion of a given ion species moves on a longer orbit and a slower ion of the same ion species moves on a shorter orbit,
such that the velocity difference is compensated by the lengths of the closed orbits~\cite{Franzke08}.
This means that the revolution times reflect directly the $m/q$ ratios of the stored ions independently of their $\Delta v/v$.
A mass resolving power of 200~000 (FWHM) has been achieved almost over the entire spectrum
with precise $B\rho$ determination at the dispersive mid-plane of the fragment separator FRS~\cite{Geissel06, Sun08},
which allows to resolve isomers with excitation energies of several hundred keV, dependend on the charge state of the isomer.

In $\gamma$-ray spectroscopy, the produced isomers are implanted in a catcher and are thus present as neutral atoms.
In a storage ring the isomeric states can be stored as fully-stripped ions,
thus providing an advantage that the internal conversion channel is disabled.
The latter is helpful for the cases where the internal conversion coefficients are experimentally unknown~\cite{Bowry13}.
Owing to dedicated time-of-flight detectors, IMS has a very high detection efficiency
and is sensitive to single stored ions~\cite{Mei10}.
Furthermore, experimental cooler storage ring (CSRe) tuned into the IMS ion-optical mode can simultaneously store ions with
a relative momentum acceptance $\Delta P/P$ of $\sim 0.2\%$~\cite{Xia02}.
The latter is important in view of the wide velocity distribution of projectile fragments.
These capabilities were employed to investigate properties of the projectile fragmentation reaction.
For instance, relative production cross sections of the fragments were measured and a relationship of the relative odd-even staggering
of yields versus the particle-emission threshold energies was established~\cite{Mei14}.

IMS can be applied to short-lived nuclides with half-lives down to $\sim$ 50 $\mu s$~\cite{Litvinov10}.
Furthermore, there is no upper half-life limitation. In this respect IMS is a complementary technique to $\gamma$-ray spectroscopy for the investigation of angular-momentum population.
This is illustrated here with the example of the measured population of the high-spin $I=19/2\hbar$ states in the mirror nuclei $^{53}$Fe and $^{53}$Co.
The half-lives of the excited states are 2.54 m and 247 ms~\cite{Audi12}, respectively.
Moreover, in the present work we investigate the relation between the isomeric ratios for a given spin and the mass loss. 
The question of how the isomeric yield ratio depends on the mass loss is not only related to the understanding of the origin of angular momentum,
but plays also an important role in the production and application of isomeric beams~\cite{Walker06} as well as spin-aligned beams~\cite{Ichikawa13}. 
In previous works ~\cite{Bacelar13, Gladnishki04, Pfutzner02, Pfutzner98, Myalski09, Steer11, Bowry13, Myalski12, Podolyak05, Podolyak06}, nuclides with the same spin had different structure and different isomer excitation energies, 
which masked a possible dependence of the isomeric ratios versus the mass loss~\cite{Gladnishki04, Daugas01}.
Thus, in order to remove the nuclear structure effects, we use the isomeric ratio values 
of the high-spin isomeric 19/2 state in $^{53}$Fe produced using different projectiles: $^{58}$Ni 
(Ref.~\cite{Irnish95}), $^{78}$Kr (present work), $^{84}$Kr (Refs.~\cite{Hausmann01, Stadlmann14}) 
and $^{112}$Sn (Ref.~\cite{Huang13}).
We note that the same method was used to study the relation between the isomeric ratios
and the mass loss in $^{44}$Sc for the case of the photo-nuclear reaction~\cite{Do08}.

\section{Experiment and Data Analysis}
The experiment was performed at the HIRFL-CSR (Cooler-Storage Ring at the Heavy Ion Research Facility in Lanzhou)~\cite{Xia02}.
The CSRe was operated in the isochronous mode
to measure masses of short-lived $T_z=-1/2$~\cite{Tu11} and $T_z=-3/2$~\cite{Zhang12} nuclei.
Many of these masses are decisive for the understanding of astrophysical nucleosynthesis and nuclear structure~\cite{Tu11, Zhang12, Tu14, Yan13}.
In this experiment, a primary $^{78}$Kr$^{28+}$ beam at an energy of 479.4~$A$ MeV was extracted
from the main storage ring (CSRm) and focused upon a $\sim$ 2.77~g/cm$^2$ beryllium production target placed
at the entrance of the radioactive beam line (RIBLL2).
$^{53}$Fe and $^{53}$Co nuclides were produced via projectile fragmentation reactions.
According to CHARGE calculations \cite{Scheidenberger-NIM1998} at this kinetic energy 
more than 99.9\% of Fe and Co fragments emerged from the target as fully-stripped nuclei.
Any contributions of other atomic charge states can be neglected in the present context.
The fragments were separated in flight with RIBLL2 and then injected into the CSRe.
To enable IMS, the CSRe was tuned to store ions at $\gamma_t\approx1.4$.
According to calculations with the LISE++ code~\cite{Tarasov04, Tarasov08},
the longitudinal momenta of $^{53}$Fe and $^{53}$Co nuclei emerging from the target are almost identical.
However, the charge state 26$^+$ for fully-ionised $^{53}$Fe is different from 27$^+$ for $^{53}$Co.
Thus, the magnetic rigidities of the RIBLL2 and CSRe facilities were sequentially set to 6.1994~Tm and 5.9708~Tm, respectively,
to center ($dP/P_c=0$) the $^{53}$Fe and $^{53}$Co fragments, respectively.
The production cross section of $^{53}$Fe is sufficiently large, so that the $^{53}$Fe nuclei with momenta
far from the central one ($dP/P_c=-3.6\%$) could be measured in the CSRe at the magnetic rigidity of 5.9708~Tm
corresponding to the central setting for $^{53}$Co.
%=================================================================
\begin{table}
\caption{Isomeric ratios (R$_{exp}$) for the 19/2 state with excitation energy
of $\sim3$~MeV in $^{53}$Fe and $^{53}$Co fragments (A$_f$) were obtained
via different projectile (A$_p$) fragmentation reactions at IMP (this work) and GSI~\cite{Irnish95, Hausmann01}.
The uncertainties of ratios only include statistical errors. 
The relative momentum settings ($dP/P_c$) are listed, where $P_c$ is the central momentum of the fragment after the target calculated by LISE++ code,
and $dP=P-P_c$ is the difference to the central momentum of the corresponding ion optical setting of the facilities.}
\begin{center}
\begin{tabular}{llllllll} \hline\hline
A$_p$ & $E$     &    ~$A_f$@dP/P$_c$      & Mass&    Spin    &     R$_{exp}$  &        Target  & Ref.        \\
 & $A$ MeV & & loss& & ~$\%$ & g/cm$^2$ & \\
\hline
$^{58}$Ni &$\sim$ 370 &  ~$^{53}$Fe $\sim$0$\%$     &     ~5  &     ~19/2    &     9.6(4)            &           4       &  ~\cite{Irnish95}   \\
$^{78}$Kr &479.4   &    ~$^{53}$Fe 0$\%$     &     ~25       &     ~19/2    &     30(4)             &           2.77    & this work        \\
$^{78}$Kr &479.4   &    ~$^{53}$Fe -3.6$\%$  &     ~25       &     ~19/2    &     38(3)             &           2.77    & this work        \\
$^{78}$Kr &486.4   &    ~$^{53}$Fe -2.1$\%$  &     ~25       &     ~19/2    &     35(4)             &           2.77    & this work        \\
$^{78}$Kr &482.9    &    ~$^{53}$Fe -4.6$\%$  &     ~25       &     ~19/2    &     39(3)*            &           2.77    & this work        \\
$^{84}$Kr &445.3    &    ~$^{53}$Fe 0$\%$     &     ~31       &     ~19/2    &     33(4)             &           2.50    & ~\cite{Hausmann01, Stadlmann14}  \\
$^{112}$Sn &395.5  &    ~$^{53}$Fe -2.9$\%$  &     ~59       &     ~19/2    &     34(2)             &           1.85    & ~\cite{Huang13}     \\
$^{78}$Kr &479.4   &    ~$^{53}$Co 0$\%$     &     ~25       &     ~19/2    &     28(14)            &           2.77    & this work        \\
$^{78}$Kr &482.9    &    ~$^{53}$Co -1$\%$    &     ~25       &     ~19/2    &     23(2)             &           2.77    & this work        \\
\hline\hline
\end{tabular}
*Ground and isomeric states were not resolved. The ratio was extracted assuming two overlapping Gaussian functions.
{\normalsize }
\label{ta3}
\end{center}
\end{table}
%=================================================================
The details of standard IMS measurements in the CSRe can be found in Refs.~\cite{Tu11-2, Mei10}.
The revolution times of the stored ions were measured by a dedicated timing detector,
see Refs.~\cite{Mei10,Zhang-NIM2014, Shuai-PLB2014} for more details on the detector design and performance.

Each stored ion passed through the thin carbon foil of the timing detector at every revolution.
Secondary electrons released from the foil due to the passage of each ion were guided to a set of micro-channel plates thus providing timing signals.
The signals from individual ions are periodic which is used to determine their revolution frequencies.
The detector efficiency ranges from 20$\%$ to 70$\%$ depending on the ion species and ion number~\cite{Mei10, Shuai-PLB2014}.
However, since each stored ion is recorded for $\sim$320 turns, a detection efficiency
of 100$\%$ can safely be assumed for all ions~\cite{Tu15}.

\begin{figure}[h!]
\begin{center}
\includegraphics[width=8.6cm]{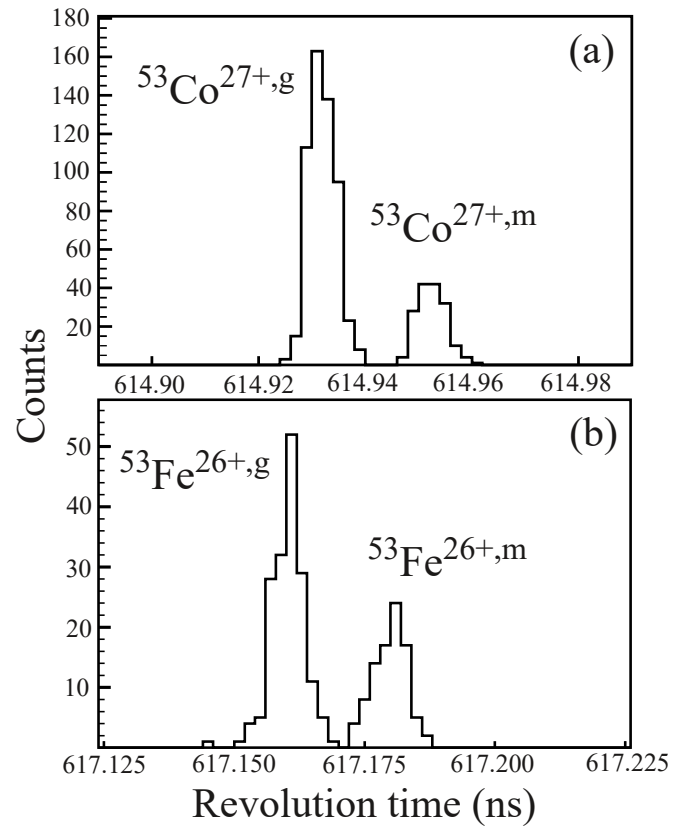}
\caption{Revolution time spectra of the isomeric and ground states of $^{53}$Co (a) and $^{53}$Fe (b) nuclides measured in the present work.
The mass difference of $\sim3$~MeV between the isomeric and ground states is reflected in a time difference of $\sim20$~ps.}
\label{fig1}
\end{center}
\end{figure}

Owing to the high mass resolving power achieved in these IMS measurements of about 170~000~\cite{Tu11-2}, the mass difference of $\sim3$~MeV~\cite{Audi12}
between the isomeric and ground states can clearly be resolved.
The revolution time spectra of the isomeric and ground states for $^{53}$Fe and $^{53}$Co are shown in Figure~\ref{fig1}.
%===============  fig. 1  ========================================
%=================================================================

The ultimate efficiency of the experimental method allows for highly accurate counting of the number of stored ions 
in the ground and isomeric states.
The experimental isomeric ratios ($R_{exp}$) are defined as the number ($N_{Jm}$) of nuclei stored in the isomeric state relative to the total
number ($N_{all}$) of stored nuclei in ground and isomeric states, as follows:
%===============  eq. 2  ========================================
\begin{equation}
R_{exp}=\frac{N_{Jm}}{N_{all}}\quad.
\label{eq11}
\end{equation}
%=================================================================

The $^{53}$Fe and $^{53}$Co nuclides were stored as bare nuclei.
The lifetimes of the isomeric and ground states are much longer than the typical measuring time of 200~$\mu$s.
Therefore, the internal conversion and $\beta$-decays during the measuring time can be neglected.
The relative mass difference of  isomeric and ground states for  $^{53}$Fe and $^{53}$Co is about $6 \times 10^{-5}$, respectively.
Thus, the transmission, injection and storage efficiencies of  isomeric and ground states are almost the same and can be neglected in the calculation of $R_{exp}$ for one magnetic rigidity setting in CSRe.

Measured isomeric yield ratios for various $dP/P_c$ settings are listed in Table~\ref{ta3}.
The available results from the experimental storage ring ESR at GSI are added as well.
The isomeric ratio of 9.6(4)$\%$ was obtained in $^{58}$Ni projectile fragmentation
at an energy around 370~$A$ MeV on a 4 g/cm$^2$ beryllium target~\cite{Irnish95}.
For the $^{58}$Ni experiment the ratio was measured for the central momentum by using SMS at the ESR of GSI~\cite{Irnish95}.
Under the condition of the thick target, the isomeric ratios with a small mass loss
are almost independent of the momentum selection~\cite{Simpson09}.
By means of the fragmentation of $^{84}$Kr projectiles at an energy of 445.3 $A$ MeV on a 2.5 g/cm$^2$ beryllium production target,
isomeric ratio of 33(4)$\%$ was measured  for $^{53}$Fe by IMS at the ESR with  relative longitudinal momentum selection $dP/P_c=0 \%$. More experimental details can be found in Ref.~\cite{Hausmann01}.
A $^{53}$Fe isomeric ratio of 34(2)$\%$ at relative longitudinal momentum selection $dP/P_c=-2.9\%$ was measured in $^{112}$Sn projectile fragmentation~\cite{Huang13}. The $^{112}$Sn beam from CSRm at an energy of 395.5 $A$ MeV impinged on a 1.85 g/cm$^2$ beryllium target placed at the entrance of RIBLL2. The $^{53}$Fe ions produced were separated and injected into CSRe by RIBLL2. CSRe was set to an isochronous condition with $\gamma_t=1.302$ and a magnetic rigidity of $B\rho = 5.306$ Tm. The revolution times for $^{53}$Fe were measured by a timing detector, as in the present work described in experiment section.

\section{RESULTS AND DISCUSSION}
In this section we address the dependence of the measured isomeric ratios for $^{53}$Fe
on the longitudinal momentum and on the mass loss.
\subsection{DEPENDENCE OF THE ISOMERIC RATIO ON THE LONGITUDINAL MOMENTUM}
\begin{figure}[h!]
\begin{center}
\includegraphics[width=8.6cm]{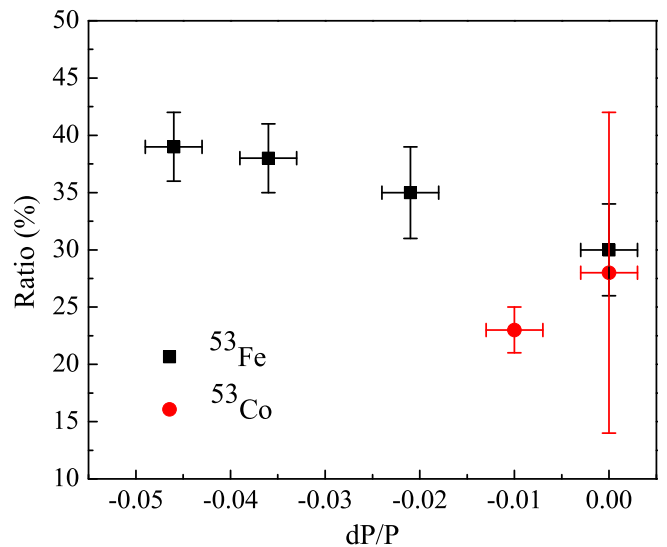}
\caption{(color online) Isomeric ratios measured in this work as a function 
of the relative longitudinal momentum selection ($dP/P=[P-P_c]/P_c$),
where, $P$ and $P_c$ are the selected momentum and the central momentum of the fragment, respectively.
Both $^{53}$Fe and $^{53}$Co are produced by $^{78}$Kr projectile fragmentation on a 2.77 g/cm$^2$ beryllium target.
}
\label{fig2}
\end{center}
\end{figure}

The dependence of the isomeric ratios on the longitudinal momentum was observed for the first time by W.-D. Schmidt-Ott et al.~\cite{Schmidt94}.
However, the dependence would be washed out for thick target and large mass loss~\cite{Daugas01, Simpson09}.
The relative momentum acceptance of the IMS at CSRe can not cover the entire
relative momentum distribution of  the fragments emerging from the target. 
Thus, the IMS can be used to study the dependence of the isomeric ratio on the momentum.
The relation between the isomeric ratios and momentum selection is also addressed in the present work.
The isomeric ratios for $^{53}$Fe produced by $^{78}$Kr projectile fragmentation with the same beam energy and the same
target thickness amount to 30(4)$\%$ for the central setting ($dP/P_c=0\%$)
and 38(3)$\%$ for the setting on the tail of the longitudinal momentum distribution ($dP/P_c=-3.6\%$).
This has to be compared to the relative momentum width of $\sim 5\%$ (FWHM) for $^{53}$Fe fragments emerging from the target.
The ratios with different momentum selections and different projectile energies were also measured, see Table~\ref{ta3}.
The results are plotted in Figure~\ref{fig2} as a function of the relative momentum difference.
The ratios of $^{53}$Fe for the central settings are slightly smaller than those corresponding to the tails of the momentum distribution.
As a conclusion, we see that there is a slight dependence of isomeric ratios on the momentum selection for $^{53}$Fe produced via $^{78}$Kr projectile fragmentation on a 2.77 g/cm$^2$ beryllium target.

%===============  fig. 2  ========================================
%=================================================================

\subsection{DEPENDENCE OF THE ISOMERIC RATIO ON THE MASS LOSS}
\begin{figure}[h!]
\begin{center}
\includegraphics[width=8.6cm]{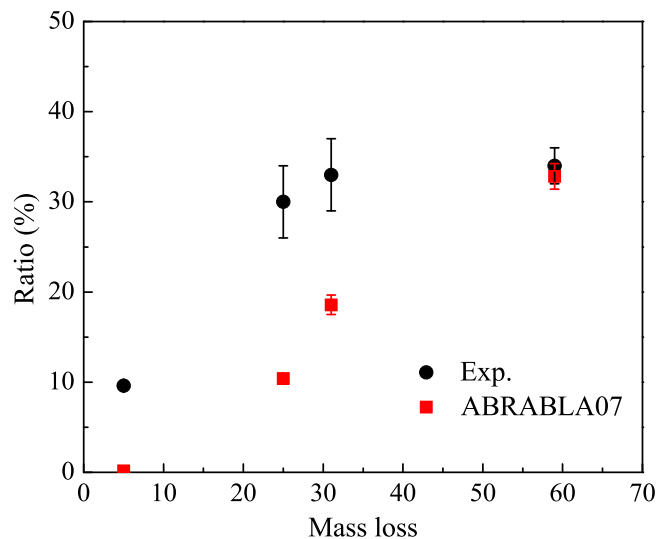}
\caption{(color online) Isomeric ratios for $^{53}$Fe as a function of the mass loss compared to the predictions of the ABRABLA07 code. The isomeric ratios rapidly increase with the projectile-fragmentation mass difference and then remain approximately constant.}
\label{fig3}
\end{center}
\end{figure}

In order to eliminate the effects of nuclear structure, the isomeric ratios of $^{53}$Fe produced 
from different projectiles, namely $^{58}$Ni~\cite{Irnish95}, $^{78}$Kr [this work], $^{84}$Kr~\cite{Hausmann01, Stadlmann14} and $^{112}$Sn~\cite{Huang13}, 
were used to study the relation between the ratios and the mass losses. 
The measured values of the isomeric ratios for $^{53}$Fe are shown in Figure~\ref{fig3} as a function of the mass loss, 
that is the difference in mass number between considered projectile and $^{53}$Fe. 
It can clearly be seen, that the value of the isomeric ratio  
increases rapidly with increasing mass loss and reaches a approximate flattop at the mass loss of about 25-30 mass units.

%===============  fig. 3  ========================================
%=================================================================

The isomeric ratios for $^{53}$Fe deduced from different projectiles as a function of the mass loss have been calculated with the two-stage abrasion-ablation code ABRABLA07~\cite{Gaimard91, Jong97, Kelic08}. 
Details on the de-excitation part of the code ABLA07 can be found in Ref.~\cite{Kelic08}. 
We assume that in the case of fragmentation reactions at high projectile energies 
the angular momentum removed by particle evaporation is small.
Thus we calculate the spin distribution of the final fragment as a superposition 
of the spins of all pre-fragments which contribute to the production of the final fragment of interest.

In Figure~\ref{fig3} we compare the experimental results with the ABRABLA07 predictions for the isomeric ratios of the spin state 19/2 in $^{53}$Fe. 
Qualitatively, the ABRABLA07 calculation describes the increasing trend of isomeric ratio with mass loss, which indicates that the population of angular momentum in a final fragment has a strong ``memory'' of the initial angular-momentum population in the pre-fragment. 
However, quantitatively, except for the largest value of the mass loss, corresponding to $^{112}$Sn projectiles, 
the calculations underestimate the isomeric ratios - the smaller the mass loss the larger is the fractional discrepancy between the data and calculations (R$_{exp}$/R$_{th}$).
The average square value of the angular momentum projection of a nucleon $<j_z^2>$ is about 2.54 for the $^{56}$Ni region in the ABRABLA07 statistical model ~\cite{Jong97}.
However, the isomer of spin state 19/2 has a three particle configuration in the case of $^{53}$Fe, each in the high angular momentum $f_{7/2}$ orbital~\cite{Eskola66}. 
Therefore the angular momentum per nucleon is much larger than the sqrt($<j_z^2>$), which may explain why the ABRABLA07 prediction is so low for $^{58}$Ni case.  

%===============  fig. 4  ========================================
\begin{figure}[h!]
\begin{center}
\includegraphics[width=8.6cm]{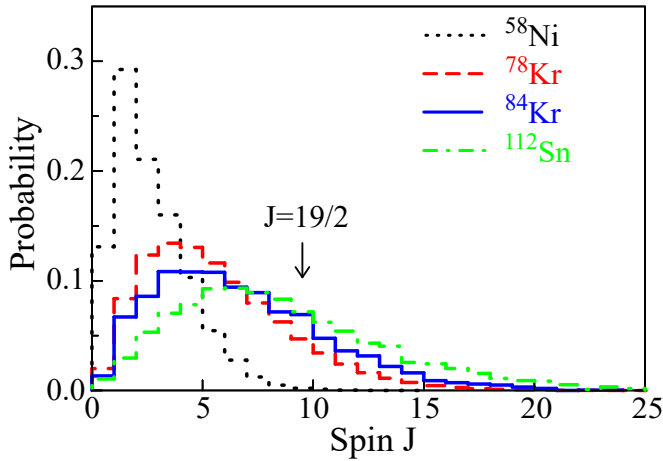}
\caption{(color online) Production probability of $^{53}$Fe versus spin calculated with ABRABLA07.
The dotted, dashed, solid and dot-dashed lines correspond to calculations performed for different projectiles 
$^{58}$Ni, $^{78}$Kr, $^{84}$Kr and $^{112}$Sn, respectively.
The spin 19/2 is indicated with an arrow.}
\label{fig4}
\end{center}
\end{figure}
%=================================================================

It has already been noticed, see, e.g., Ref.~\cite{Podolyak06}, 
that for high-spin states the values of isomeric ratios calculated with the ABRABLA code are lower than the measured ones. 
This is consistent with the present case, since the probability to produce higher spin states depends on the mass loss. 
Figure~\ref{fig4} illustrates the probabilities to produce $^{53}$Fe with different spins for different projectiles calculated with ABRABLA07 code.
It can be seen that the spin 19/2 can hardly be produced for a small mass loss ($^{58}$Ni projectiles).
However, it is easily produced for larger mass losses (Kr and Sn projectiles).
The observation of this work suggests that the calculated underestimation (R$_{exp}$/R$_{th}$) of the isomeric ratios
depends not only on the spin~\cite{Podolyak06}, but also on the mass loss.

%===============  fig. 5  ========================================
\begin{figure}[h!]
\begin{center}
\includegraphics[width=8.6cm]{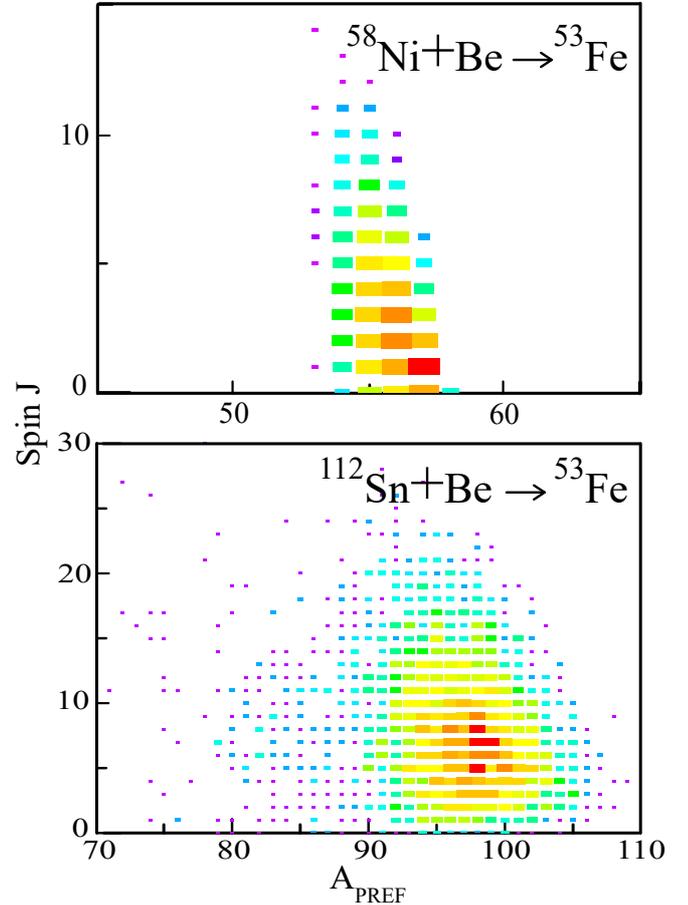}
\caption{(color online) Angular momentum versus mass of the pre-fragments leading to the observed $^{53}$Fe 
final fragment produced in the fragmentation of $^{58}$Ni and $^{112}$Sn calculated with ABRABLA07 code.}
\label{fig5}
\end{center}
\end{figure}
%=================================================================

Several effects have been discussed as possible origins of the discrepancy between 
calculated and measured isomeric ratios for high-spin states~\cite{Podolyak06,Pal08,Bowry13}. 
Although a clear answer concerning the population of different spin states is still missing, 
data shown in Figure~\ref{fig3} can serve as a benchmark for further developments in theoretical models. 
Fluctuations in angular momentum, mass and nuclear charge distributions of pre-fragments produced in 
the fragmentation of $^{112}$Sn leading to $^{53}$Fe final fragments may be large enough to properly reproduce observed isomeric ratios. 
These fluctuations ``overwrite'' any other effect influencing angular momentum in the fragmentation reactions. 
However, for small mass losses these fluctuations are also small, see Figure~\ref{fig5}, since the number 
of different pre-fragments leading to the final fragment is rather small.
These are thus the cases where additional contributions to angular-momentum population have to be investigated.

\section{Summary}
The Isochronous Mass Spectrometry (IMS) method was applied to 
investigate angular momentum populations of high-spin isomers in projectile fragmentation reactions.
Isomeric ratios were measured for the spin 19/2 state in the mirror  nuclei $^{53}$Fe and $^{53}$Co
following the projectile fragmentation of $^{78}$Kr with energies of $\sim$480~$A$ MeV on a beryllium target at HIRFL-CSR.
The half-lives of the excited states of $^{53}$Fe and $^{53}$Co are 2.54~m and 247~ms [39], respectively,
in which cases the IMS can be seen as a complementary technique to the $\gamma$-ray spectroscopy
for the study of angular momentum populations, especially for systems with long lifetimes.
The relationship of the isomeric ratio and momentum selection was studied by measured isomeric ratios of $^{53}$Fe.
The experimental isomeric ratios for the same spin have been extracted as a function of the mass loss.
Data obtained in this work and from literature show that the isomeric ratios rapidly increase with the mass loss and then saturate.
This increasing behavior is approximately reproduced by the ABRABLA07 calculations. 
Quantitatively, the calculations can reproduce data for the 19/2 spin state in $^{53}$Fe for the largest mass loss, 
but considerably underestimate the corresponding data for small mass losses.
Our new data point to a need for further developments of the theoretical description of the fragmentation reaction process.
The latter is indispensable for the planning of experiments at operating facilities as well as for the next-generation facilities like FAIR or HIAF.

%\section*{Acknowledgements}
This work is supported in part by
NSFC (grant Nos. 11205205, 11575112, U1232208, 11135005, 11605252, 11605267 and 11605249), 
by the CAS Pioneer Hundred Talents Program,
by the 973 Program of China (No. 2013CB 834401), 
by the BMBF grant in the framework of the Internationale Zusammenarbeit
in Bildung und Forschung Projekt-Nr. 01DO12012, by the Helmholtz-CAS
Joint Research Group HCJRG-108, by the External Cooperation
Program of the Chinese Academy Sciences Grant No. GJHZ1305, by the UK STFC, by the Max-Plank-Society,
and by the European Research Council (ERC) under the European Union's Horizon 2020 research and innovation programme (grant agreement No
682841 ``ASTRUm'').

\end{document}